\documentclass
[hyperref,aps,amsfonts,amssymb,showpacs,showkeys,preprint,letterpaper,12pt]{revtex4}%
\usepackage{amsfonts}
\usepackage{amsmath}
\usepackage{amssymb}
\usepackage{graphicx}%
\providecommand{\U}[1]{\protect\rule{.1in}{.1in}}
\newtheorem{theorem}{Theorem}
\newtheorem{acknowledgement}[theorem]{Acknowledgement}

\begin{document}
\preprint{LPTh/001-22}
\title[Particle creation and minimal length]{Particle creation in the presence of minimal length: The time dependent gauge}
\author{M. Bouali}
\affiliation{\textit{LPTh, Department of Physics, University of Jijel, BP 98, Ouled Aissa,
Jijel 18000, Algeria.}}
\author{S. Haouat}
\affiliation{\textit{LPTh, Department of Physics, University of Jijel, BP 98, Ouled Aissa,
Jijel 18000, Algeria.}}

\begin{abstract}
In this paper we have studied the problem of scalar particles pair creation by
a constant electric field in the presence of a minimal length. A closed
expression for the corresponding Green's function is obtained via path
integral approach. Then by projecting this function on the outgoing particle
and antiparticle states we have calculated the probability to create a pair of
particles and the number density of created particles. From this, we have
deduced the modifications brought by the minimal length to Hawking temperature
and black hole entropy. It is shown that the first correction is a logarithmic
term with a negative numerical factor. We have also examined the semiclassical
WKB approximation in the calculation of the pair production rate. The result
is that, unlike the ordinary case, the WKB approximation in the presence of a
minimal length does not give the exact rate even for the constant electric field.

\end{abstract}
\date{}
\maketitle

\section{Introduction}

The existence of a minimal measurable length is a common feature to many
phenomenological approaches to quantum gravity \cite{Sabine,Tawfik}. In 1964,
Mead has shown, through gedanken experiments, that the gravitational
interactions make it impossible to measure the position of a particle with an
uncertainty less than \cite{Mead}
\[
l_{P}=\sqrt{\frac{G\hbar}{c^{3}}}\simeq1.61605\times10^{-35}%
\operatorname{m}%
\]
where $c$ is the speed of light in vacuum, $\hbar$ is the Planck constant and
$G$ is the gravitational constant (the Newton constant). This fundamental
scale existed, indeed, long before the theory of general relativity; in 1899,
Planck used the universal constants $c$, $\hbar$ and $G$ to form, amongst
others, a fundamental unit of length, which refers to as the Planck length.
Nevertheless, at that date the Planck length, despite its universality, was
not immediately admitted as an important scale. After the theory of general
relativity was established, this fundamental length has found its
significance. It can be understood as the scale at which the Schwarzschild
radius of a black hole is equal to its Compton wavelength and therefore, the
scale at which both gravitational and quantum effects become significant.
Having taken on this quite meaning, the Planck length has become the
indispensable ingredient in many phenomenological approaches to quantum
gravity \cite{GUP1,GUP2,GUP3,GUP4}. It arises in string theory
\cite{ST1,ST2,ST3,ST4}, loop quantum gravity \cite{LQG}, black hole physics
\cite{BH1,BH2,BH3} and in non-commutative field theories \cite{NC1,NC2,NC3}.

The existence of such a minimal length in nature would change to some extent
our understanding of short distance physics and hence very high energy
physics. This explains why various physical problems are reconsidered by
taking into account the minimal length either at quantum or classical level
\cite{Casadio,Scardigli}. Besides the large number of simple quantum
mechanical systems \cite{MQ1,MQ2,MQ3,MQ4,MQ5,MQ6,MQ7,MQ8,MQ9}, such as the
harmonic oscillator and the Hydrogen atom, the effect of minimal length have
been the subject of many studies . As example, we cite the influence of the
minimal length on the Casimir effect \cite{Casimir1,Casimir2}, the Unruh
effect \cite{Unruh1}, the Hawking radiation and black hole thermodynamics
\cite{BHT1,BHT2,BHT3,BHT4,BHT5,BHT6} and the Schwinger effect
\cite{HK,BRMu,SCML}. Elsewhere, the quantum gravity corrections to the mean
field theory of nucleons have been, recently, communicated \cite{ANaqash}.

In this paper, we propose to study the particle-antiparticle pair creation
from vacuum by an electric field in the presence of a minimal length. This
effect, which refers to as the Schwinger effect, has been predicted in quantum
electrodynamics several decades ago \cite{EH,Schwinger,Gelis}. It can be
interpreted as follows: In Minkowski space-time, the free quantum field has an
invariant vacuum state under Poincar\'{e} symmetry, which defines a unique set
of creation and annihilation operators and thence makes possible the
interpretation of the field in terms of particles and antiparticles. This,
together with relativistic covariance, implies that all inertial observers
agree on the same number of particles contained in a quantum state. In the
presence of an external electric field, the vacuum state is no longer
invariant under space-time translations and, therefore, it is no longer
possible to define a state that would be identified by all observers as the
vacuum state. Because of this, the interpretation of the quantum field in
terms of particles might be ambiguous since, in such a case, the vacuum is
instable. In particular, the particle creation occurs spontaneously when the
electric field breaks the invariance under space-time translations, implying
instantaneous vacuum state. In other words, the state initially recognized as
the vacuum state evolves in time and could contain outgoing particles at a
very distant time.

On the other hand, the minimal length is, generally, implemented in the
quantum theory through a generalized uncertainty principle (GUP) which implies
a deformation of the Poincar\'{e} group. Hence, in order to formulate a
quantum field theory with a minimal length, it is necessary to carefully
retrace all steps of the standard quantization scheme \cite{Hossenfelder}.
However, to our knowledge, there is no well-established quantum field theory
that consistently includes such a minimal length. It is thus of great interest
to examine the accuracy of usual methods in the study of several physical
processes in the presence of minimal length. This includes the Unruh effect,
the Hawking radiation, the particle creation and many other phenomena. In this
regard, the Schwinger effect is a good example to clarify several issues. It
could give a clear answer to the following questions:

\begin{itemize}
\item When a minimal length is present, is the Schwinger effect still
consistent with gauge invariance principle? In Schwinger effect, particles and
antiparticles are produced in pairs, i.e. for each created particle an
antiparticle is created simultaneously. Consequently, it is expected that the
Schwinger effect preserves the electric charge conservation and, thus, the
gauge invariance. A constant electric field can be described by the space
dependent gauge, where $A_{0}=-eEx$, or by a time-dependent gauge;
$A_{z}=-eEt$. In the ordinary case, without minimal length, we know how the
creation of particles in these two gauges are related to each other
\cite{Pad}. However, in the presence of a minimal length, it is not clear how
the two gauges give the same results.

\item The second question is about the validity of the semiclassical
approximation: It is well-known that the use of the semiclassical WKB
approximation reproduce the exact pair production rate by constant electric
field in Minkowski space-time. Indeed, the sum of all contributions from
higher order corrections cancels. In the presence of a minimal length existing
studies shows that the semiclassical approximation does not give the same
results for the pair creation rate as the exact treatment. See for instance
\cite{HK} and \cite{SCML}.

\item How the Schwinger effect is connected to the Hawking radiation and Black
hole entropy? The black hole radiation in the presence of minimal length and
corrected black hole thermodynamics have been much written about, and remain
attracting much attentions. The most used method in this context, besides some
heuristic derivations, is the semiclassical WKB method, which has been
considered \ as the method yielding more accurate results. However, since the
accuracy of this method on the study of the Schwinger effect is called into
question, it would be useful to discuss the possible link between Schwinger
effect and Hawking radiation. The comparison between the two effects could
lead us to many important findings.
\end{itemize}

We notice that in reference \cite{HK}, that one of us was involved in, the
creation of scalar particles by an electric field in the presence of a minimal
length is investigated. In that work, the authors considered a charged scalar
particle subjected to a constant electric field expressed in space dependent
gauge. Then, since the corresponding Klein Gordon equation is exactly soluble
in momentum space, they were able to calculate the pair production probability
and the number density of created particles by the use of the Bogoliubov
transformation connecting the "in" with the "out" states. Recently, the
particle creation in the presence of minimal length is reconsidered in the
context of WKB approximation for a deformed Schrodinger-like equation
\cite{SCML}. Otherwise, the problem has been discussed in different scenarios
such as GUP models involving minimal length and/or maximal momentum
\cite{Ong,Hamil1,Hamil2,BRMu} and noncommutative geometry \cite{NCPC1,NCPC2}.
This is because the Schwinger effect has become an interdisciplinary research
area due to its considerable number of applications in various fields of
physics, from heavy nuclei to black hole physics \cite{Ruffini}. In this
paper, in order to satisfactorily answer the questions evoked above, we
reconsider the Schwinger effect in the presence of a minimum length by using
the time-dependent gauge.

The rest of this paper proceeds as follows. In Sec. 2, we provide some
arguments that lead to estimate heuristically the effect of the minimal length
on the creation of particles. In Sec. 3, we introduce a generalized Heisenberg
uncertainty principle to derive the Klein Gordon equation with minimal length
for a charged particle in an electric field. In Sec. 4, we use the path
integral method to calculate the corresponding Green's function. In Sec. 5, we
calculate the probability to create one pair of particles. In Sec. 6, we
discuss the application of the Schwinger effect with minimal length in black
holes radiation. In Sec. 7, we examine the validity of semiclassical
WKB\ approximation. Sec. 8 is reserved for concluding remarks.

\section{The minimal length and particle creation}

In ordinary quantum electrodynamics, it has been shown that the pair
production rate in constant electric field of strength $E$ is given by
\cite{CF1}
\begin{equation}
\mathcal{W}\sim\exp\left(  -\pi\frac{m^{2}c^{3}}{eE\hbar}\right)  , \label{w1}%
\end{equation}
what confirms the nonperturbative nature of the process and stipulates a
critical value above which the effect becomes appreciable%
\begin{equation}
E_{c}=\frac{m^{2}c^{3}}{e\hbar}\sim10^{16}\text{V/cm}.
\end{equation}
In this section, we try to find an adequate interpretation to the above
formula by considering some intuitive arguments, which could be generalized
straightforwardly to the case when a minimal length exists.

Let us first recall that the creation and annihilation of particle pairs, is
related to the difficulties encountered when unifying quantum mechanics and
special relativity in a single-particle theory. In relativistic quantum
mechanics, it is the Heisenberg's uncertainty principle that leads to the
creation of particles. Indeed, when a particle is confined to a region of size
$\Delta x$, it has, according to Heisenberg's uncertainty relation, a momentum
width $\Delta p\geq\frac{\hbar}{2}\frac{1}{\Delta x}$. Consequently, if
$\Delta x\lesssim\lambda_{C}$, with $\lambda_{C}=\frac{\hbar}{mc}$, the
particle's momentum uncertainty becomes%
\begin{equation}
\Delta p\geq\left(  \Delta p\right)  _{0}=\frac{1}{2}mc, \label{dp0}%
\end{equation}
where $\left(  \Delta p\right)  _{0}=\frac{1}{2}mc$ is the smaller momentum
uncertainty that accompanies the localization of a mass $m$ particle with
precision equal to its Compton wavelength. Therefore, the energy uncertainty
of a relativistic particle becomes of the same order as its rest mass,
$\Delta\mathcal{E}\sim\Delta pc\sim mc^{2}$. This gives rise to the creation
of virtual particle-antiparticle pairs. However, since the lifetime of these
pairs is very short, they finish by conversely annihilating.

Nevertheless, when a virtual pair is subjected to a constant electric field of
strength $E$, the particle and its antiparticle move apart from each other and
they will gain the energy $eEL$ after covering a relative distance $L$. If the
gained energy exceeds the rest mass of the two particles, $eEL\geq2mc^{2}$,
the pair will become real and the particles will continue to move apart.
However, since the typical separation of the virtual pair is of order of the
Compton wavelength, $L\sim\frac{\hbar}{mc}$, it follows that%
\begin{equation}
eE\gtrsim\frac{m^{2}c^{3}}{\hbar}.
\end{equation}
This explains way there is a critical value from which the effect becomes susceptible.

The aim now is to seek an intuitive interpretation for the pair production
rate. To this aim, we first write the exponent in equation (\ref{w1}) in the form%

\begin{equation}
\pi\frac{m^{2}c^{3}}{eE\hbar}=\pi\frac{mc^{2}}{eE\times\frac{\hbar}{mc}}.
\end{equation}
This can be viewed as the ratio of the particle rest energy to the electric
field energy over a distance equal to the particle Compton wavelength
\cite{Aitchison}. At first glance, this observation does not seem in any way
generalizable to the case of minimal length. Instead, if we write%
\[
\pi\frac{m^{2}c^{3}}{eE\hbar}=2\pi\frac{mc^{2}}{eE}\times\frac{mc}{2}%
=2\pi\frac{l\times\left(  \Delta p\right)  _{0}}{\hbar}%
\]
where $\left(  \Delta p\right)  _{0}$ is given by equation (\ref{dp0}) and
$l=\frac{mc^{2}}{eE}$ is the minimal distance travelled by each particle of
the pair in the electric field to reach the energy $2mc^{2}$, we obtain%
\begin{equation}
\mathcal{W}\sim\exp\left(  -2\pi\frac{l\times\left(  \Delta p\right)  _{0}%
}{\hbar}\right)  ,
\end{equation}
which can be immediately extended to case of minimal length. To show this, we
consider the GUP
\begin{equation}
\Delta P\Delta X\geq\frac{\hbar}{2}\left[  1+\beta\left(  \Delta P\right)
^{2}+...\right]  , \label{eq:GUP}%
\end{equation}
where $\beta$ is very small positive parameter. Such a GUP leads to a nonzero
minimal length given by
\begin{equation}
\left(  \Delta X\right)  _{\min}=\hbar\sqrt{\beta}. \label{eq:MinLength}%
\end{equation}
From (\ref{eq:GUP}), we can see that for a particle localized with precision
$\Delta X$, we have%
\begin{equation}
\Delta P\geq\frac{\Delta X}{\beta\hbar}\left(  1-\sqrt{1-\frac{\beta\hbar^{2}%
}{\left(  \Delta X\right)  ^{2}}}\right)  ,
\end{equation}
$\allowbreak$ Therefore, if $\Delta X$ is smaller than the Compton wave length
of the particle, i.e. $\Delta X\leq\frac{\hbar}{mc}$, the uncertainty on
momentum becomes
\[
\Delta P\geq\left(  \Delta P\right)  _{0}=\frac{1}{\beta mc}\left(
1-\sqrt{1-\beta m^{2}c^{2}}\right)
\]
As in ordinary case (without minimal length), we can guess the probability of
particle creation to be%
\begin{equation}
\mathcal{W}=\exp\left(  -2\pi\frac{l\times\left(  \Delta P\right)  _{0}}%
{\hbar}\right)  =\exp\left[  -2\pi\frac{c}{eE\beta\hbar}\left(  1-\sqrt
{1-\beta m^{2}c^{2}}\right)  \right]
\end{equation}
This result, although heuristic handwaving, is exactly the same as that
obtained by semi-classical method \cite{SCML}. However, by taking only the
correction to the first order
\begin{equation}
\mathcal{W}=\exp\left[  -2\pi\frac{c}{eE\beta\hbar}\left(  1-\sqrt{1-\beta
m^{2}c^{2}}\right)  \right]  \simeq\exp\left[  -\pi\frac{m^{2}c^{3}}{eE\hbar
}\left(  1+\frac{\beta}{4}m^{2}c^{2}+...\right)  \right]  .
\end{equation}
we can see that this result differs from that of the reference \cite{HK},
which is believed to have been obtained by an exact approach,
\begin{equation}
\mathcal{W}=\exp\left[  -\pi\frac{m^{2}c^{3}}{eE\hbar}\left(  1+\frac{1}%
{4}\beta m^{2}c^{2}\left(  1-\frac{e^{2}E^{2}\hbar^{2}}{m^{4}c^{6}}\right)
\right)  \right]  . \label{w2}%
\end{equation}
In light of the above review which emphasizes clearly the discrepancy between
the mentioned results, the following remarks are in order. The first one is
that the authors of \cite{HK} expressed the Klein Gordon equation in the
momentum space representation,
\begin{equation}
\left[  \left(  \omega+eE\hat{X}\right)  ^{2}-\hat{P}^{2}-m^{2}\right]
\varphi=0, \label{40}%
\end{equation}
where the action of the operators $\hat{X}$ and $\hat{P}$ is defined by%
\begin{align}
\hat{P}  &  =p,\label{eq:momentum}\\
\hat{X}  &  =i\hbar\left[  \left(  1+\beta p^{2}\right)  \frac{\partial
}{\partial p}\right]  . \label{eq:momentum2}%
\end{align}
Besides this momentum representation, we find for the operators $\hat{X}$ and
$\hat{P}$ many other representations. However, it is not yet known if these
representations are equivalent and how they are related to one another.

Secondly, we notice that in both \cite{HK} and \cite{SCML}, the authors
consider a space dependent gauge. Actually, it is not clear whether the use of
time dependent gauge will give the same results. This question should be quite
investigated because any result for the pair creation rate does not make sense
as long as it does not satisfy the gauge invariance principle.

In response to these questions it is of great interest to reconsider the
problem by using a different approach. In this context, one can consider a
position representation for the $\hat{P}$ operator and a time dependent gauge
for the electric field. One can look also for a method beyond the Bogoliubov
transformation and the WKB approximation. This is the main goal of rest of the paper.

In what follows, we shall use the system of units with $c=\hbar=1$ but we
shall show explicitly $G$. In addition, for the sake of simplicity, we shall
consider the Schwinger effect in (1+1) dimensional space-time. It is
well-known that, apart from the unimportant prefactor, the pair production
rate does not depend on the space-time dimension.

\section{The Klein Gordon equation with minimal length}

In the ordinary quantum mechanics, the momentum and the wave vector are
related to one another by the linear relation $p=k$ (note that $\hbar$ is
taken to be $1$). This means that the physical moment is the same as the
canonical moment. In order to reproduce the GUP in (\ref{eq:GUP}), we consider
a nonlinear relation of the form $p=f(k)$, where $f$ is an injective function
so that $f^{-1}(p)=k$ is well defined \cite{Sabine}. This requires a
redefinition of physical operators in terms of canonical operators. The
physical operators $\hat{X}$ and $\hat{P}$ are now defined by
\begin{align}
\hat{X}  &  =\hat{x}\label{x}\\
\hat{P}  &  =f\left(  \hat{p}\right)  , \label{p}%
\end{align}
where $\hat{x}$ and $\hat{p}$ are the canonical operators obeying the usual
commutation relation $\left[  \hat{x},\hat{p}\right]  =i$. Of course, there
are infinitely many functions $f$ that could give rise to a minimal length.
The simpler case is to consider the expansion%
\begin{equation}
f\left(  \hat{p}\right)  =\hat{p}\left(  1+\frac{\beta}{3}\hat{p}%
^{2}+...\right)  . \label{fp}%
\end{equation}
Taking into account that $\left[  \hat{x},f\left(  \hat{p}\right)  \right]
=if^{\prime}\left(  \hat{p}\right)  $ and $\tilde{p}\approx\hat{P}\left(
1-\frac{\beta}{3}\hat{P}^{2}+...\right)  $, we obtain the modified commutation
relation
\begin{equation}
\left[  \hat{X},\hat{P}\right]  =i\left(  1+\beta\hat{P}^{2}\right)  ,
\label{CR1}%
\end{equation}
that leads to equation (\ref{eq:GUP}).

From equations (\ref{x}), (\ref{p}) and (\ref{fp}), we can define for the
operators $\hat{X}$ and $\hat{P}$ the following position space representation%
\begin{align}
\hat{X}  &  =x,\\
\hat{P}  &  =\left(  1+\frac{1}{3}\beta\hat{p}^{2}\right)  \hat{p},
\end{align}
with%
\begin{equation}
\ \hat{p}=-i\frac{\partial}{\partial x},
\end{equation}
In the position space representation, the free Klein Gordon equation to the
first order in $\beta$ is given by%
\begin{equation}
\left(  -\frac{\partial^{2}}{\partial t^{2}}+\frac{\partial^{2}}{\partial
x^{2}}-\frac{2\beta}{3}\frac{\partial^{4}}{\partial x^{4}}-m^{2}\right)
\psi\left(  t,x\right)  =0, \label{free}%
\end{equation}
With this formulation that includes the gravitational minimal length, it is
straightforward to show that the obtained Klein Gordon equation is consistent
with the gauge invariance. In effect, assuming that a charged scalar particle
couples to the electromagnetic field following the minimal coupling principle
which consists on making the change
\begin{align}
i\frac{\partial}{\partial t}  &  \rightarrow i\frac{\partial}{\partial
t}-eA_{0}\\
\text{\ }i\frac{\partial}{\partial x}  &  \rightarrow i\frac{\partial
}{\partial x}-eA_{1},
\end{align}
we obtain, all to the first order in $\beta$, the following Klein Gordon
equation%
\begin{equation}
\left[  \left(  i\frac{\partial}{\partial t}-eA_{0}\right)  ^{2}-\left(
i\frac{\partial}{\partial x}-eA_{1}\right)  ^{2}-\frac{2\beta}{3}\left(
i\frac{\partial}{\partial x}-eA_{1}\right)  ^{4}-m^{2}\right]  \psi\left(
t,x\right)  =0. \label{gke}%
\end{equation}
This equation is invariant under the gauge field transformation%
\begin{equation}
A_{\mu}^{\prime}\left(  t,x\right)  =A_{\mu}\left(  t,x\right)  +\partial
_{\mu}\chi,
\end{equation}
with%
\[
\psi^{\prime}\left(  t,x\right)  =e^{ie\alpha}\psi\left(  t,x\right)
\]
Let us, now, consider a scalar particle of mass $m$ and charge $e$ subjected
to a constant electric field $E$. In the presence of a minimal length the use
of the time-dependent gauge, with the assumption that $\psi\left(  t,x\right)
=e^{-ipx}\varphi\left(  t\right)  $, leads to the following differential
equation%
\begin{equation}
\left[  \frac{\partial^{2}}{\partial t^{2}}+\left(  p+eEt\right)  ^{2}%
+\frac{2\beta}{3}\left(  p+eEt\right)  ^{4}+m^{2}\right]  \varphi\left(
t\right)  =0. \label{tgke}%
\end{equation}
To our knowledge, because of the quartic term $\left(  p+eEt\right)  ^{4}$,
there is no exact solution for this differential equation. However, as is
shown in \cite{Cheriet}, one can find approximate solutions (to the first
order on $\beta$) by making the change $t\rightarrow\xi$ with%
\begin{equation}
\xi=\frac{1}{2}-\sqrt{\frac{\beta}{12}}\left(  eEt+p\right)  .
\end{equation}
The resulting equation can be put in the form of hypergeometric equation
\begin{equation}
\left[  \xi\left(  1-\xi\right)  \frac{\partial^{2}}{\partial\xi^{2}}+\left(
C-\left(  A+B+1\right)  \xi\right)  \allowbreak\frac{\partial}{\partial\xi
}-AB\right]  F\left(  \xi\right)  =0. \label{hypeq}%
\end{equation}
where the constants $A$, $B$ and $C$ are given by
\begin{align}
A  &  =\frac{1}{2}+i\frac{1}{\alpha}\sqrt{1-\lambda\alpha-\alpha^{2}}+\frac
{i}{\alpha}\sqrt{1-2\lambda\alpha-\frac{1}{4}\alpha^{2}}\\
B  &  =\frac{1}{2}+i\frac{1}{\alpha}\sqrt{1-\lambda\alpha-\alpha^{2}}-\frac
{i}{\alpha}\sqrt{1-2\lambda\alpha-\frac{1}{4}\alpha^{2}}\\
C  &  =1+i\frac{1}{\alpha}\sqrt{1-\lambda\alpha-\alpha^{2}}%
\end{align}
with%
\begin{align}
\alpha &  =\frac{\beta}{3}eE,\\
\lambda &  =\frac{m^{2}}{eE}.
\end{align}
The solution of the original equation is related to the new solution by
\begin{equation}
\varphi\left(  t\right)  =\xi^{a}\left(  1-\xi\right)  ^{b}F\left(
\xi\right)
\end{equation}
where $a$ and $b$ are given by
\begin{equation}
a=b=\frac{1}{2}+i\frac{1}{2\alpha}\sqrt{1-\lambda\alpha-\alpha^{2}}.
\end{equation}
The two independent solutions for the hypergeometric equation (\ref{hypeq})
are given by \cite{Grad}%
\begin{align}
F_{1}\left(  \xi\right)   &  =F\left(  A,B;C;\xi\right) \\
F_{2}\left(  \xi\right)   &  =\xi^{1-C}F\left(  A-C+1,B-C+1;2-C;\xi\right)
\end{align}
Then, the outgoing particle state is given by \cite{Cheriet}%
\begin{equation}
\varphi_{out}^{+}\left(  t\right)  =N\xi^{\alpha_{1}}\left(  1-\xi\right)
^{\alpha_{1}}F\left(  2\alpha_{1}+\alpha_{2},2\alpha_{1}-\alpha_{2}%
-1,2\alpha_{1};\xi\right)  \label{kiout}%
\end{equation}
with the normalization constant%
\begin{equation}
N=2\left(  \frac{\alpha}{eE\left(  1-\lambda\alpha-\alpha^{2}\right)
}\right)  ^{\frac{1}{4}}.
\end{equation}
Having shown how to investigate the Klein Gordon equation in the presence of a
minimal length, we are now able to study the creation of particle pairs.
Because of the nonperturbative nature of this effect, several practical
techniques have been developed in order to calculate the pair production rate.
Among these methods we cite the Schwinger effective action method
\cite{EA1,EA2,EA3,EA4} and related instantons calculus
\cite{Ins1,Ins2,Ins3,Ins4}, the Hamiltonian diagonalization technique
\cite{Grib,Grib1}, the "in" and "out" states formalism
\cite{Che1,Che2,Che3,Che4}, the quantum kinetic approach \cite{qka1,qka2,qka3}
as well as the method based on the projection of Green's function on the
particle states \cite{GF1,GF2,GF3,GF4,GF5,GF6} and the semiclassical WKB
approximation \cite{Semcla1,Semcla2,Semcla3}.

In this work, we consider the Green's function method. This method has been
proved to be most fruitful in finding the probability to create a pair of
particles either in external electromagnetic fields or in cosmological backgrounds.

We start with the calculation of the corresponding Green's function using
Feynman path integrals.

\section{Path integral derivation of the Green's function}

In the presence of a minimal length and the propagator of a relativistic
particle subjected to an electric field described by the gauge $A^{\mu
}=\left(  A^{0},A_{x}\right)  =\left(  0,-eEt\right)  $ is the causal Green's
function solution of the equation
\begin{equation}
\hat{O}_{KG}G\left(  x_{f},x_{i},t_{f},t_{i}\right)  =\delta\left(
x_{f}-x_{i}\right)  \delta\left(  t_{f}-t_{i}\right)
\end{equation}
where $\hat{O}_{KG}$ is the Klein Gordon operator%

\begin{equation}
\hat{O}_{KG}=\left(  \hat{p}^{0}\right)  ^{2}-\left(  \hat{p}+eEt\right)
^{2}-\frac{2}{3}\beta\left(  \hat{p}+eEt\right)  ^{4}-m^{2},
\end{equation}
where $\hat{p}^{0}=i\frac{\partial}{\partial t}$ and $\hat{p}=-i\frac
{\partial}{\partial x}$. It is well-known that $G\left(  x_{f},x_{i}%
,t_{f},t_{i}\right)  $ can be represented as a matrix element of an operator
$F^{-1}$
\begin{equation}
G\left(  x_{f},x_{i},t_{f},t_{i}\right)  =\left\langle x_{f},t_{f}\right\vert
F^{-1}\left\vert x_{i},t_{i}\right\rangle
\end{equation}
where the operator $F$ is given by
\begin{equation}
F=-\left(  \hat{p}^{0}\right)  ^{2}+\left(  \hat{p}+eE\hat{t}\right)
^{2}+\frac{2}{3}\beta\left(  \hat{p}+eE\hat{t}\right)  ^{4}+m^{2}%
\end{equation}
and $\left\vert x,t\right\rangle $ are eigenvectors for the operators $\hat
{t}$ and $\hat{x}$;
\[
\hat{x}\left\vert x,t\right\rangle =x\left\vert x,t\right\rangle
~\ ;~\ \hat{t}\left\vert x,t\right\rangle =t\left\vert x,t\right\rangle
\]
For the corresponding canonical-conjugated operators of momenta $\hat{p}^{0}$
and $\hat{p}$, we have
\[
\hat{p}^{0}\left\vert p,p^{0}\right\rangle =p^{0}\left\vert p,p^{0}%
\right\rangle ~\ ;~\ \hat{p}\left\vert p,p^{0}\right\rangle =p\left\vert
p,p^{0}\right\rangle
\]
Since these operators satisfy the usual commutation relations $\left[  \hat
{x},\hat{p}\right]  =i$ and $\left[  \hat{t},\hat{p}^{0}\right]  =i$, we have
for the eigenvectors the usual normalization and completeness relations
\[
\int dxdt\left\vert x,t\right\rangle \left\langle x,t\right\vert
=1~~\ ;~\ \left\langle x,t\right.  \left\vert x^{\prime},t^{\prime
}\right\rangle =\delta\left(  x-x^{\prime}\right)  \delta\left(  t-t^{\prime
}\right)
\]
and%
\[
\int dp^{0}dp\left\vert p,p^{0}\right\rangle \left\langle p,p^{0}\right\vert
=1~~\ ;~\left\langle p,p^{0}\right.  \left\vert p^{\prime},p^{\prime
0}\right\rangle =\delta\left(  p^{0}-p^{\prime0}\right)  \delta\left(
p-p^{\prime}\right)  .
\]
We define also a plane wave with the scalar product%
\begin{equation}
\left\langle p,p^{0}\right.  \left\vert x,t\right\rangle =\frac{1}{2\pi
}e^{-i\left(  p^{0}t-px\right)  }. \label{pw}%
\end{equation}
By the use of the integral representation%
\begin{equation}
F^{-1}=i\int_{0}^{\infty}dT\exp\left[  -i\left(  F-i\epsilon\right)  T\right]
\end{equation}
where $T$ is the Schwinger proper-time and $i\epsilon$ is an infinitesimal
imaginary number that can be included to $m^{2}$ and has to be put to zero at
the end of calculations, we obtain the proper time representation of the Green
function $G\left(  x_{f},,x_{i},t_{f},,t_{i}\right)  $%

\begin{equation}
G\left(  x_{f},,x_{i},t_{f},,t_{i}\right)  =i\overset{\infty}{\underset
{0}{\int}}dT~K\left(  x_{f},,x_{i};t_{f},,t_{i};T\right)  , \label{D}%
\end{equation}
where the kernel$~K\left(  x_{f},,x_{i};t_{f},,t_{i};T\right)  $ is given by%
\begin{equation}
K\left(  x_{f},,x_{i};t_{f},,t_{i};T\right)  =\left\langle x_{f}%
,t_{f}\right\vert \exp\left(  -iFT\right)  \left\vert x_{i},t_{i}\right\rangle
.
\end{equation}
Here $K\left(  x_{f},,x_{i};t_{f},,t_{i};T\right)  $ is similar to the
quantum-mechanical amplitude for the transition between an initial state
$\left\vert x_{i},t_{i}\right\rangle $ at the proper time $0$ and the final
state $\left\vert x_{f},t_{f}\right\rangle ~$at proper time $T$. To present
$K\left(  x_{f},,x_{i};t_{f},,t_{i};T\right)  $ by means of path integrals we
first write $\exp\left(  -iFT\right)  =\left[  \exp\left(  -iF\varepsilon
\right)  \right]  ^{N+1},$ with $\varepsilon=1/\left(  N+1\right)  $, and we
insert $N$ identities $\int dxdt\left\vert x,t\right\rangle \left\langle
x,t\right\vert =1$ between all the operators $\exp\left(  -iF\varepsilon
\right)  $. We obtain%
\begin{align}
K\left(  x_{f},,x_{i};t_{f},,t_{i};T\right)   &  =\int dx_{1}dx_{2}%
~...dx_{N}\int dt_{1}dt_{2}~...dt_{N}\nonumber\\
&  \times\prod\limits_{n=1}^{N+1}\left\langle x_{n},t_{n}\right\vert
\exp\left(  -iF\varepsilon\right)  \left\vert x_{n-1},t_{n-1}\right\rangle .
\label{13}%
\end{align}
Now, we have to express the matrix elements $\left\langle x_{n},t_{n}%
\right\vert \exp\left(  -iF\varepsilon\right)  \left\vert x_{n-1}%
,t_{n-1}\right\rangle $ through path integral. As $\varepsilon$ is small, we
can write%
\begin{equation}
\left\langle x_{n},t_{n}\right\vert \exp\left(  -iF\varepsilon\right)
\left\vert x_{n-1},t_{n-1}\right\rangle \approx\left\langle x_{n}%
,t_{n}\right\vert \left(  1-i\varepsilon F\right)  \left\vert x_{n-1}%
,t_{n-1}\right\rangle . \label{14}%
\end{equation}
Then, we insert in (\ref{14}) the integral identity $\int dp_{n}^{0}%
dp_{n}\left\vert p_{n},p_{n}^{0}\right\rangle \left\langle p_{n},p_{n}%
^{0}\right\vert =1$. By taking into account that $F$ has no product of the
noncommutating operators $\hat{p}^{0}$ and $\hat{t}$, and by using the plane
wave definition (\ref{pw}) the matrix element (\ref{14}) can be expressed in
the middle point $\bar{t}_{n}=(t_{n}+t_{n-1})/2$ as follows%
\begin{align}
&  \left.  \left\langle x_{n},t_{n}\right\vert \exp\left(  -iF\varepsilon
\right)  \left\vert x_{n-1},t_{n-1}\right\rangle =\right. \nonumber\\
&  \int\frac{dp_{n}^{0}dp_{n}}{\left(  2\pi\right)  ^{2}}\exp\left\{  i\left[
p_{n}^{0}~\left(  \triangle t_{n}\right)  -p_{n}~\left(  \triangle
x_{n}\right)  \right.  \right. \nonumber\\
&  \left.  \left.  -\varepsilon\left(  \left(  p_{n}^{0}\right)  ^{2}-\left(
p+eE\bar{t}_{n}\right)  ^{2}-\frac{2}{3}\beta\left(  p_{n}+eE\bar{t}%
_{n}~\right)  ^{4}-m^{2}\right)  \right]  \right\}  . \label{16}%
\end{align}
where $\Delta q_{n}=q_{n}-q_{n-1}$ for $q\equiv x,t$. We have then%

\begin{align}
K\left(  x_{f},,x_{i};t_{f},,t_{i};T\right)   &  =\int\prod_{n=1}^{N}%
dx_{n}\prod_{n=1}^{N}dt_{n}\prod_{n=1}^{N+1}\frac{dp_{n}}{2\pi}\prod
_{n=1}^{N+1}\frac{d\left(  p_{0}\right)  _{n}}{2\pi}\nonumber\\
&  \times\prod_{n=1}^{N+1}\exp\left\{  i\left[  p_{n}^{0}~\left(  \triangle
t_{n}\right)  -p_{n}~\left(  \triangle x_{n}\right)  -\varepsilon\left(
p_{n}^{0}\right)  ^{2}\right.  \right. \nonumber\\
&  \left.  \left.  +\varepsilon\left(  \left(  p_{n}+eE\bar{t}_{n}\right)
^{2}+\frac{2}{3}\beta\left(  p_{n}+eE\bar{t}_{n}~\right)  ^{4}+m^{2}\right)
\right]  \right\}  ,
\end{align}
By doing integrations over $p_{n}^{0}$, $p_{n}$ and $x_{n}$, we obtain%
\begin{equation}
K\left(  x_{f},,x_{i};t_{f},,t_{i};T\right)  =\int\frac{dp}{2\pi}e^{-ip\left(
x_{f}-x_{i}\right)  }\mathcal{K}_{p}\left(  t_{f},t_{i};T\right)  \label{S.}%
\end{equation}
where the kernel $\mathcal{K}_{p}\left(  t_{f},t_{i};T\right)  $ is given by%
\begin{align}
\mathcal{K}_{p}\left(  t_{f},t_{i};T\right)   &  =\int\underset{n=1}%
{\overset{N}{\prod}}dt_{n}\underset{n=1}{\overset{N+1}{\prod}}\sqrt{\frac
{1}{4i\pi\varepsilon}}\nonumber\\
&  \exp\left\{  i\varepsilon\sum_{n=1}^{N+1}\left[  \frac{\left(  \triangle
t_{n}\right)  ^{2}}{4\varepsilon^{2}}+\left(  p+eE\bar{t}_{n}\right)
^{2}+\frac{2}{3}\beta\left(  p+eE\bar{t}_{n}\right)  ^{4}+m^{2}\right]
\right\}  . \label{kp}%
\end{align}
Then the Green's function $G\left(  x_{f},x_{i},t_{f},t_{i}\right)  $ can be
written as%

\begin{equation}
G\left(  x_{f},x_{i},t_{f},t_{i}\right)  =\int\frac{dp}{2\pi}e^{-ip\left(
x_{f}-x_{i}\right)  }\mathcal{G}_{p}\left(  t_{f},t_{i}\right)
\end{equation}
with%
\begin{equation}
\mathcal{G}_{p}\left(  t_{f},t_{i}\right)  =i\overset{\infty}{\underset
{0}{\int}}dT~\mathcal{K}_{p}\left(  t_{f},t_{i};T\right)  .
\end{equation}
The aim now is to calculate $\mathcal{G}_{p}\left(  t_{f},t_{i}\right)  $, for
this purpose we write $\mathcal{K}_{p}\left(  t_{f},t_{i};T\right)  $ in its
standard form%

\begin{equation}
\mathcal{K}_{p}\left(  t_{f},t_{i};T\right)  =\int Dt\exp\left[  i\int_{0}%
^{T}d\tau\left[  \frac{\dot{t}^{2}}{4}+\left(  p+eEt\right)  ^{2}+\frac{2}%
{3}\beta\left(  p+eEt\right)  ^{4}+m^{2}\right]  \right]  \label{pi1}%
\end{equation}
with the measure%
\begin{equation}
Dt=\underset{n=1}{\overset{N}{\prod}}dt_{n}\underset{n=1}{\overset{N+1}{\prod
}}\sqrt{\frac{1}{4i\pi\varepsilon}}%
\end{equation}
It is well-known that, due to the quartic term $\left(  p+eEt\right)  ^{4}$,
the path integral in (\ref{pi1}) has no exact solution. The use of
approximations is therefore absolutely necessary. This is possible as long as
$\beta$ is a small parameter. Let us first introduce the transformation%

\begin{equation}
d\tilde{\tau}=\frac{d\tau}{\rho^{2}\left(  t\right)  }%
\end{equation}
with%
\begin{equation}
\rho\left(  t\right)  =1-\frac{\beta}{3}\left(  p+eEt\right)  ^{2}.
\end{equation}
This transformation eliminates the term with $\left(  p+eEt\right)  ^{4}$ and
makes hence the problem approximately soluble. To show this we have first to
express the slicing parameter $\varepsilon$ in terms of the new parameter
$\tilde{\varepsilon}_{n}$ defined by
\begin{equation}
\tilde{\varepsilon}_{n}=\tilde{\tau}_{n}-\tilde{\tau}_{n-1}=\int_{\tau_{n-1}%
}^{\tau_{n}}\frac{d\tau}{\rho^{2}\left(  t\right)  }.
\end{equation}
Developing the last integral as%
\begin{equation}
\tilde{\varepsilon}_{n}=\frac{\Delta\tau}{\rho^{2}\left(  t_{n-1}\right)
}\left[  1-\frac{\dot{\rho}\left(  t_{n-1}\right)  }{\rho\left(
t_{n-1}\right)  }\left(  \Delta\tau\right)  \right]
\end{equation}
and writing
\[
1-\frac{\dot{\rho}\left(  t_{n-1}\right)  }{\rho\left(  t_{n-1}\right)
}\left(  \Delta\tau\right)  \simeq\exp\left(  -\frac{\dot{\rho}\left(
t_{n-1}\right)  }{\rho\left(  t_{n-1}\right)  }\left(  \Delta\tau\right)
\right)  =\exp\left(  -\int_{\tau_{n-1}}^{\tau_{n}}\frac{\dot{\rho}\left(
t\right)  }{\rho\left(  t\right)  }d\tau\right)  ,
\]
we obtain
\begin{equation}
\tilde{\varepsilon}_{n}=\frac{\Delta\tau}{\rho^{2}\left(  t_{n-1}\right)
}\exp\left(  -\ln\frac{\rho\left(  t_{n}\right)  }{\rho\left(  t_{n-1}\right)
}\right)  =\frac{\varepsilon}{\rho\left(  t_{n}\right)  \rho\left(
t_{n-1}\right)  }.
\end{equation}
Next, by taking into account that
\begin{equation}
\underset{n=1}{\overset{N+1}{\prod}}\sqrt{\frac{1}{\rho\left(  t_{n}\right)
\rho\left(  t_{n-1}\right)  }}=\frac{1}{\sqrt{\rho\left(  t_{f}\right)
\rho\left(  t_{i}\right)  }}\underset{n=1}{\overset{N}{\prod}}\frac{1}%
{\rho\left(  t_{n}\right)  },
\end{equation}
we can rearrange the kernel $\mathcal{K}_{p}\left(  t_{f},t_{i};T\right)  $ as
follows%
\begin{align}
\mathcal{K}_{p}\left(  t_{f},t_{i};T\right)   &  =\sqrt{\frac{1}{\rho\left(
t_{f}\right)  \rho\left(  t_{i}\right)  }}\int\underset{n=1}{\overset{N}%
{\prod}}\frac{dt_{n}}{\rho\left(  t_{n}\right)  }\underset{n=1}{\overset
{N+1}{\prod}}\sqrt{\frac{1}{4i\pi\tilde{\varepsilon}_{n}}}\nonumber\\
&  \exp\left\{  i\sum_{n=1}^{N+1}\left[  \frac{\left(  \Delta t_{n}\right)
^{2}}{4\tilde{\varepsilon}_{n}\rho\left(  t_{n}\right)  \rho\left(
t_{n-1}\right)  }+\tilde{\varepsilon}_{n}\left(  \allowbreak\left(  1-\frac
{2}{3}\beta m^{2}\right)  \left(  p+eEt_{n}\right)  ^{2}+m^{2}\right)
\right]  \right\}  .
\end{align}
The transformation from $\tau$ to $\tilde{\tau}$, implies a change on the
proper time from $T$ to $S$ with
\begin{equation}
T=\int_{S_{i}}^{S_{f}}\rho^{2}\left(  t\right)  d\tau
\end{equation}
and $S=S_{f}-S_{i}$. Therefore, in order to incorporate the novel proper time
we use the identity%
\begin{equation}
\rho\left(  t_{f}\right)  \rho\left(  t_{i}\right)  \int_{0}^{\infty}%
dS\delta\left(  T-\int_{0}^{S}\rho^{2}\left(  t\right)  d\tau\right)  =1.
\end{equation}
As a result, we obtain for the Green's function $\mathcal{G}_{p}\left(
t_{f},t_{i}\right)  $%
\begin{align}
\mathcal{G}_{p}\left(  t_{f},t_{i}\right)   &  =i\sqrt{\rho\left(
t_{f}\right)  \rho\left(  t_{i}\right)  }\overset{\infty}{\underset{0}{\int}%
}dS~\int\underset{n=1}{\overset{N}{\prod}}\frac{dt_{n}}{\rho\left(
t_{n}\right)  }\underset{n=1}{\overset{N+1}{\prod}}\sqrt{\frac{1}{4i\pi
\tilde{\varepsilon}_{n}}}\nonumber\\
&  \exp\left\{  i\sum_{n=1}^{N+1}\left[  \frac{\left(  \Delta t_{n}\right)
^{2}}{4\tilde{\varepsilon}_{n}\rho\left(  t_{n}\right)  \rho\left(
t_{n-1}\right)  }+\tilde{\varepsilon}_{n}\left(  \allowbreak\left(  1-\frac
{2}{3}\beta m^{2}\right)  \left(  p+eEt_{n}\right)  ^{2}+m^{2}\right)
\right]  \right\}
\end{align}
Here we remark an inconvenient time dependence in both the measure
$\frac{dt_{n}}{\rho\left(  t_{n}\right)  }$ and the kinetic term
$\frac{\left(  \Delta t_{n}\right)  ^{2}}{4\tilde{\varepsilon}_{n}\rho\left(
t_{n}\right)  \rho\left(  t_{n-1}\right)  }$. It is then necessary to
eliminate this dependence by introducing the transformation $t_{n}\rightarrow
u_{n}$, with
\begin{equation}
du_{n}=\frac{dt_{n}}{\rho\left(  t_{n}\right)  }%
\end{equation}
or, equivalently,%
\begin{equation}
t_{n}+\frac{p}{eE}=\frac{1}{eE\sqrt{\frac{\beta}{3}}}\tanh\left(
eE\sqrt{\frac{\beta}{3}}u_{n}\right)  .
\end{equation}
In this case we have
\begin{equation}
\frac{\left(  \Delta t_{n}\right)  ^{2}}{\rho\left(  t_{n}\right)  \rho\left(
t_{n-1}\right)  }=\left(  \Delta u_{n}\right)  ^{2}+\frac{1}{9}e^{2}E^{2}%
\beta\left(  \Delta u_{n}\right)  ^{4}+...
\end{equation}
and consequently, the Green's function $\mathcal{G}_{p}\left(  t_{f}%
,t_{i}\right)  $ will be given by
\begin{align}
\mathcal{G}_{p}\left(  t_{f},t_{i}\right)   &  =i\sqrt{\rho\left(
t_{f}\right)  \rho\left(  t_{i}\right)  }\overset{\infty}{\underset{0}{\int}%
}dS~\int\underset{n=1}{\overset{N}{\prod}}du_{n}\underset{n=1}{\overset
{N+1}{\prod}}\sqrt{\frac{1}{4i\pi\tilde{\varepsilon}_{n}}}\nonumber\\
&  \exp\left\{  i\sum_{n=1}^{N+1}\left[  \frac{1}{4\tilde{\varepsilon}_{n}%
}\left(  \left(  \Delta u_{n}\right)  ^{2}+\frac{1}{9}e^{2}E^{2}\beta\left(
\Delta u_{n}\right)  ^{4}\right)  \right.  \right. \nonumber\\
&  \left.  \left.  +\tilde{\varepsilon}_{n}\left(  \left(  \frac{3}{\beta
}-2m^{2}\right)  \tanh^{2}\left(  eE\sqrt{\frac{\beta}{3}}\bar{u}_{n}\right)
+m^{2}\right)  \right]  \right\}  . \label{gpnc}%
\end{align}
According to McLaughlin-Schulman procedure \cite{McLaughlin}, the sum
$i\sum\limits_{n=1}^{N+1}\frac{1}{9}e^{2}E^{2}\beta\frac{\left(  \Delta
u_{n}\right)  ^{4}}{4\tilde{\varepsilon}_{n}}$ can be replaced by a sum of the
form $i\sum\limits_{n=1}^{N+1}\tilde{\varepsilon}_{n}Q_{n}$, where $Q_{n}$ can
be calculated with the help of the property%
\begin{equation}
\overset{\infty}{\underset{-\infty}{\int}}\exp\left(  -ax^{2}-bx^{4}\right)
dx=\overset{\infty}{\underset{-\infty}{\int}}\exp\left(  -ax^{2}-\frac
{3b}{4a^{2}}\right)  dx+\mathcal{O}\left(  \frac{1}{a^{3}}\right)  .
\end{equation}
This means that the term with $\left(  \Delta u_{n}\right)  ^{4}$ leads to a
quantum correction of the form
\begin{equation}
\frac{i}{4\tilde{\varepsilon}_{n}}\frac{e^{2}E^{2}\beta}{9}\left\langle
\left(  \Delta u_{n}\right)  ^{4}\right\rangle =-i\tilde{\varepsilon}_{n}%
\frac{1}{3}e^{2}E^{2}\beta.
\end{equation}
By incorporating this correction in the path integral (\ref{gpnc}) and by
taking into account that%
\begin{equation}
\tanh^{2}\left(  eE\sqrt{\frac{\beta}{3}}u_{n}\right)  =1-\frac{1}{\cosh
^{2}\left(  eE\sqrt{\frac{\beta}{3}}u_{n}\right)  }%
\end{equation}
we obtain the path integral%
\begin{align}
&  \left.  \mathcal{G}_{p}\left(  t_{f},t_{i}\right)  =i\sqrt{\rho\left(
t_{f}\right)  \rho\left(  t_{i}\right)  }\overset{\infty}{\underset{0}{\int}%
}dS~\int\underset{n=1}{\overset{N}{\prod}}du_{n}\underset{n=1}{\overset
{N+1}{\prod}}\sqrt{\frac{1}{4i\pi\tilde{\varepsilon}_{n}}}\right. \nonumber\\
&  \exp\left\{  i\sum_{n=1}^{N+1}\left[  \frac{1}{4\tilde{\varepsilon}}\left(
\Delta u_{n}\right)  ^{2}-\tilde{\varepsilon}_{n}\left(  \frac{\left(
\frac{3}{\beta}-2m^{2}\right)  }{\cosh^{2}\left(  eE\sqrt{\frac{\beta}{3}}%
\bar{u}_{n}\right)  }-\frac{3}{\beta}+m^{2}+\frac{1}{3}\beta e^{2}%
E^{2}\right)  \right]  \right\}
\end{align}
$\allowbreak$which can be put in the well-known form%
\begin{align}
\mathcal{G}_{p}\left(  t_{f},t_{i}\right)   &  =i\sqrt{\frac{3}{\beta
e^{2}E^{2}}}\sqrt{\rho\left(  t_{f}\right)  \rho\left(  t_{i}\right)
}\nonumber\\
&  \int_{0}^{+\infty}d\tilde{T}\ \int D\xi\exp\left\{  i\int_{0}^{\tilde{T}%
}d\sigma\left[  \frac{\dot{\xi}^{2}}{4}-m_{1}^{2}+\frac{l\left(  l+1\right)
}{\cosh^{2}\xi}\right]  \right\}
\end{align}
by making the rescaling
\begin{align*}
eE\sqrt{\frac{\beta}{3}}u  &  \rightarrow\xi\\
d\tilde{\tau}  &  \rightarrow\frac{3}{\beta e^{2}E^{2}}d\sigma\\
\tilde{T}  &  \rightarrow\frac{3}{\beta e^{2}E^{2}}S.
\end{align*}
The parameters $m_{1}$ and $l$ are given by%
\begin{align}
m_{1}  &  =\frac{i}{\alpha}\sqrt{1-\lambda\alpha-\alpha^{2}}=i\mu\label{mu}\\
l  &  =-\frac{1}{2}+\frac{i}{\alpha}\sqrt{1-2\lambda\alpha-\frac{1}{4}%
\alpha^{2}}=-\frac{1}{2}+i\nu. \label{nu}%
\end{align}
The problem is then reduced to a quantum mechanical system with a particular
case of the generalized Poschl-Teller potential \cite{Grosch} whose solution
is presented by
\begin{align}
\mathcal{G}_{p}\left(  t_{f},t_{i}\right)   &  =i2\sqrt{\frac{3}{\beta
e^{2}E^{2}}}\sqrt{\rho\left(  t_{f}\right)  \rho\left(  t_{i}\right)  }%
\frac{\Gamma\left(  m_{1}-l\right)  \Gamma\left(  m_{1}+l+1\right)  }%
{\Gamma\left(  m_{1}+1\right)  \Gamma\left(  m_{1}+1\right)  }\nonumber\\
&  \left(  \frac{1}{4}-\frac{\beta}{12}\left(  p+eEt_{f}\right)  ^{2}\right)
^{\frac{1}{2}+\frac{m_{1}}{2}}\left(  \frac{1}{4}-\frac{\beta}{12}\left(
p+eEt_{i}\right)  ^{2}\right)  ^{\frac{1}{2}+\frac{m_{1}}{2}}\nonumber\\
&  F\left(  m_{1}-l,m_{1}+l+1,m_{1}+1;\frac{1}{2}+\sqrt{\frac{\beta}{12}%
}\left(  p+eEt_{f}\right)  \right) \nonumber\\
&  F\left(  m_{1}-l,m_{1}+l+1,m_{1}+1;\frac{1}{2}-\sqrt{\frac{\beta}{12}%
}\left(  p+eEt_{i}\right)  \right)  .
\end{align}
This is an approximate but rigorous path integral derivation of the causal
Green's function associated with the Klein Gordon equation with minimal length
in the presence of a constant electric field, described by the time dependent gauge.

\section{Pair creation amplitude}

The one-pair creation amplitude $A\left(  p_{f},p_{i}\right)  $ is obtained by
projecting the causal Green's function on an outgoing particle and
anti-particle states. We then have%
\begin{equation}
A\left(  p_{f},p_{i}\right)  =A_{0}\int dx_{f}\int dx_{i}\psi_{p_{f}%
,out}^{+\ast}\left(  x_{f},t_{f}\right)  \chi_{out}^{+\ast}\left(
t_{f}\right)  \overleftrightarrow{\partial_{t_{f}}}G\left(  x_{f},t_{f}%
;x_{i},t_{i}\right)  \overleftrightarrow{\partial_{t_{i}}}\psi_{p_{i}%
,out}^{+\ast}\left(  x_{i},t_{i}\right)
\end{equation}
where $A_{0}$ is the amplitude for no particle production and the derivative
$\overleftrightarrow{\partial_{t}}$ is defined by
\begin{equation}
f\overleftrightarrow{\partial_{t}}g=f\frac{\partial}{\partial t}%
g-g\frac{\partial}{\partial t}f
\end{equation}
The outgoing particle state is given by
\begin{equation}
\psi_{p,out}^{+}\left(  x,t\right)  =\frac{1}{\sqrt{2\pi}}e^{ipx}%
\varphi_{p,out}^{+}\left(  t\right)  .
\end{equation}
By doing integration over $x_{i}$ and $x_{f}$, we obtain%

\begin{equation}
A\left(  p_{f},p_{i}\right)  =\delta\left(  p_{f}+p_{i}\right)  S\left(
p_{i}\right)  \label{App}%
\end{equation}
where $S\left(  p\right)  $ is given by%
\begin{equation}
S\left(  p\right)  =A_{0}\varphi_{p,out}^{+\ast}\left(  t_{f}\right)
\overleftrightarrow{\partial_{t_{f}}}\mathcal{G}_{p}\left(  t_{f}%
,t_{i}\right)  \overleftrightarrow{\partial_{t_{i}}}\varphi_{p,out}^{+\ast
}\left(  t_{i}\right)  .
\end{equation}
The delta function appearing in equation (\ref{App}) is due to the spatial
part of the outgoing states. It shows that the particle and its anti-particle
are created with opposite momenta. However, when this delta function is
squared, it leads to an inconvenient divergence of the form $\delta\left(
0\right)  $. Usually this divergence can be eliminated by employing the
following trick%
\begin{equation}
\left[  \delta\left(  p_{f}+p_{i}\right)  \right]  ^{2}=\lim_{L\rightarrow
\infty}\int_{-\frac{L}{2}}^{+\frac{L}{2}}\frac{dx}{2\pi}e^{-ix\left(
p_{f}+p_{i}\right)  }\delta\left(  p_{f}+p_{i}\right)  =\lim_{L\rightarrow
\infty}\frac{L}{2\pi}\delta\left(  p_{f}+p_{i}\right)
\end{equation}
We have then the total probability to create a pair of particles%
\begin{equation}
\int\left\vert A\left(  p_{f},p_{i}\right)  \right\vert ^{2}dp_{f}dp_{i}=\int
dp_{i}\frac{L}{2\pi}\left\vert S\left(  p_{i}\right)  \right\vert ^{2}%
=\int\frac{dxdp}{2\pi}\left\vert S\left(  p\right)  \right\vert ^{2}%
\end{equation}
Since $\frac{dxdp}{2\pi}$ is the number of states in the volume element $dxdp$
of the phase space, the quantity $\left\vert S\left(  p\right)  \right\vert
^{2}$ is therefore interpreted as the probability to create one pair in the
state $p$.

To calculate $S\left(  p\right)  $, we take into account that the particle and
its antiparticle are created simultaneously at a given time $t$. Then we must
take the limit $t_{f}\rightarrow t_{i}\rightarrow t$ after doing derivations.
In addition, outgoing particles are well-defined only at very late times after
any transient behavior has disappeared. Being aware of this condition and
taking into account that $\beta$ is so small, we take $t$ so that
\begin{equation}
p+eEt\sim\sqrt{\frac{3}{\beta}}.
\end{equation}
Besides the fact that this limit facilitates the calculation, it is consistent
with the general case of particle creation by a time dependent source, in
which particle states are determined at very late time where the source
vanishes. In effect, if we write the Klein Gordon equation in the form
\begin{equation}
\left(  \frac{\partial^{2}}{\partial t^{2}}+\omega^{2}\right)  \chi\left(
t\right)  =0,
\end{equation}
with
\begin{equation}
\omega^{2}=\left(  p+eEt\right)  ^{2}+\frac{2\beta}{3}\left(  p+eEt\right)
^{4}+m^{2},
\end{equation}
we can see that, in the limit $p+eEt\sim\sqrt{\frac{3}{\beta}}$,
\begin{equation}
\frac{\dot{\omega}}{\omega^{2}}\sim eE\beta\left(  1-\frac{1}{6}\beta
m^{2}\right)  ,
\end{equation}
which implies that the adiabatic condition is verified and consequently the
particles are well-defined.

In this limit the state $\varphi_{p,out}^{+}\left(  t\right)  $ behaves like
\begin{equation}
\varphi_{p,out}^{+}\left(  t\right)  =\frac{N}{2}e^{-\frac{m_{1}}{2}\ln
2}\left(  1-\sqrt{\frac{\beta}{3}}\left(  p+eEt\right)  \right)  ^{\frac{1}%
{2}+\frac{m_{1}}{2}}%
\end{equation}
and the Green's function $\mathcal{G}_{p}\left(  t_{f},t_{i}\right)  $ takes
the form%
\begin{align}
\mathcal{G}_{p}\left(  t_{f},t_{i}\right)   &  =ie^{-\frac{m_{1}}{2}\ln2}%
\sqrt{\frac{3}{\beta e^{2}E^{2}}}\nonumber\\
&  \frac{\Gamma\left(  m_{1}-l\right)  \Gamma\left(  m_{1}+l+1\right)
}{\Gamma\left(  m_{1}+1\right)  \Gamma\left(  m_{1}+1\right)  }F\left(
m_{1}-l,m_{1}+l+1,m_{1}+1;1\right) \nonumber\\
&  \left(  1-\sqrt{\frac{\beta}{3}}\left(  p+eEt_{f}\right)  \right)
^{\frac{1}{2}+\frac{m_{1}}{2}}\times\left(  1-\sqrt{\frac{\beta}{3}}\left(
p+eEt_{i}\right)  \right)  ^{\frac{1}{2}+\frac{m_{1}}{2}}.
\end{align}
Then the pair creation amplitude$\allowbreak$ $S\left(  p\right)  $ can be
easily calculated. The result is%
\[
S\left(  p\right)  =A_{0}m_{1}\frac{\Gamma\left(  m_{1}-l\right)
\Gamma\left(  m_{1}+l+1\right)  }{\Gamma\left(  m_{1}+1\right)  \Gamma\left(
m_{1}+1\right)  }F\left(  m_{1}-l,m_{1}+l+1,m_{1}+1;1\right)  .
\]
By taking into account that%
\[
F\left(  A,B;C;1\right)  =\frac{\Gamma\left(  C\right)  \Gamma\left(
C-A-B\right)  }{\Gamma\left(  C-A\right)  \Gamma\left(  C-B\right)  }%
\]
and%
\[
\Gamma\left(  m_{1}+1\right)  =m_{1}\Gamma\left(  m_{1}\right)  ,
\]
we obtain%
\[
S\left(  p\right)  =A_{0}\frac{\Gamma\left(  -m_{1}\right)  }{\Gamma\left(
m_{1}\right)  }\frac{\Gamma\left(  m_{1}-l\right)  \Gamma\left(
m_{1}+l+1\right)  }{\Gamma\left(  l+1\right)  \Gamma\left(  -l\right)  }.
\]
and consequently,%
\[
\left\vert S\left(  p\right)  \right\vert ^{2}=\left\vert A_{0}\right\vert
^{2}\left\vert \frac{\Gamma\left(  \frac{1}{2}+i\mu-i\nu\right)  \Gamma\left(
\frac{1}{2}+i\mu+i\nu\right)  }{\Gamma\left(  \frac{1}{2}+i\nu\right)
\Gamma\left(  \frac{1}{2}-i\nu\right)  }\right\vert ^{2}%
\]
where $\mu$ and $\nu$ are defined in (\ref{mu}) and (\ref{nu}). By taking into
account that%
\begin{equation}
\left\vert \Gamma(\frac{1}{2}+ix)\right\vert ^{2}=\frac{\pi}{\cosh\pi x}
\label{gamma}%
\end{equation}
we obtain%
\[
\left\vert S\left(  p\right)  \right\vert ^{2}=\left\vert A_{0}\right\vert
^{2}\frac{\cosh^{2}\pi\nu}{\cosh\pi\left(  \mu+\nu\right)  \cosh\pi\left(
\mu-\nu\right)  }.
\]
which can be written as%
\[
\left\vert S\left(  p\right)  \right\vert ^{2}=\left\vert A_{0}\right\vert
^{2}\frac{1}{e^{\pi\Lambda}+1}%
\]
where%
\[
e^{\pi\Lambda}=\frac{\sinh^{2}\left(  \pi\mu\right)  }{\cosh^{2}\left(  \pi
\nu\right)  }.
\]
We now have to determine the constant $A_{0}$. To this aim, we define the
probability to create $n$ pairs of particles in the state $p$ by
\[
P_{n}=~\left\vert A_{0}\right\vert ^{2}\mathcal{P}^{n}%
\]
where $\mathcal{P}$ is defined by
\[
\mathcal{P}~=\frac{1}{e^{\pi\Lambda}+1}.
\]
We then have%
\[
\sum\limits_{n=0}^{\infty}P_{n}=\sum\limits_{n=0}^{\infty}\mathcal{P}%
^{n}~\left\vert A_{0}\right\vert ^{2}=\frac{\left\vert A_{0}\right\vert ^{2}%
}{1-\mathcal{P}}=1
\]
which implies that $\left\vert A_{0}\right\vert ^{2}$ is given by
\begin{equation}
\left\vert A_{0}\right\vert ^{2}=\frac{1}{1+e^{-\pi\Lambda}} \label{3.53}%
\end{equation}
Another important result is that the average number of particles created in
the state $p$ is%
\begin{align}
\overline{n}  &  =\sum\limits_{n=0}^{\infty}nP_{n}~\label{3.58}\\
&  =\left\vert A_{0}\right\vert ^{2}\sum\limits_{n=1}^{\infty}n\mathcal{P}%
^{n}\\
&  =\left\vert A_{0}\right\vert ^{2}\frac{\mathcal{P}}{\left(  1-\mathcal{P}%
\right)  ^{2}},
\end{align}
It follows that%
\begin{equation}
\overline{n}=e^{-\pi\Lambda}=\frac{\cosh^{2}\left(  \pi\nu\right)  }{\sinh
^{2}\left(  \pi\mu\right)  },
\end{equation}
Let us, now, calculate the total number of created particles by doing
summation over all states. The total number of created particles is given by%
\begin{equation}
N=\int\frac{dxdp}{2\pi}e^{-\pi\Lambda}. \label{Ntot}%
\end{equation}
Here we note that the minimal length as introduced in (\ref{CR1}) has no
influence on the number of states $\frac{dxdp}{2\pi}$ because the time and
energy operators in the Klein Gordon equation satisfy the ordinary canonical
commutation relation.

Equation (\ref{Ntot}) can in turn be related to the rate of the Schwinger pair
production. Just replace the integration over $p$ by%
\begin{equation}
\int dp=eE\int dt.
\end{equation}
The total number of created particles is then given by%
\begin{equation}
N=\int dtdx\frac{eE}{2\pi}e^{-\pi\Lambda}. \label{1000}%
\end{equation}
On the other hand, if we write $N$ as%
\begin{equation}
N=\int dN=\int\frac{dN}{dtdx}dtdx, \label{1001}%
\end{equation}
we interpret $\frac{dN}{dtdx}\equiv\mathcal{N}$ as the number of created
particles per unit of time per unit of length. It follows from equations
(\ref{1000}) and (\ref{1001}) that
\begin{equation}
\mathcal{N}=\frac{dN}{dtdx}=\frac{eE}{2\pi}e^{-\pi\Lambda}.
\end{equation}
Since $\beta$ is small, the constants $\mu$ and $\nu$ are large and
\begin{equation}
e^{-\pi\Lambda}=\frac{\cosh^{2}\left(  \pi\nu\right)  }{\sinh^{2}\left(
\pi\mu\right)  }\simeq\frac{\exp\left(  2\pi\nu\right)  }{\exp\left(  2\pi
\mu\right)  }=e^{-2\pi\left(  \mu-\nu\right)  },
\end{equation}
what implies that
\begin{equation}
\Lambda=2\left(  \mu-\nu\right)
\end{equation}
Thus, to the first order on $\alpha$, we have%
\begin{equation}
\Lambda=\lambda+\alpha\left(  \frac{3}{8}\lambda^{2}-\frac{3}{8}\right)
=\frac{m^{2}}{eE}\left[  1+\frac{\beta}{4}\left(  m^{2}-\frac{e^{2}E^{2}%
}{m^{2}}\right)  \right]
\end{equation}
and consequently, the number of created particles per unit of time per unit of
length is finally
\begin{equation}
\mathcal{N}=\frac{eE}{2\pi}\exp\left[  -\pi\frac{m^{2}}{eE}\left(  1+\frac
{1}{4}\beta m^{2}\left(  1-\frac{e^{2}E^{2}}{m^{4}}\right)  \right)  \right]
. \label{appn}%
\end{equation}
This is exactly equation (96) of \cite{HK}.

In comparison with \cite{HK}, we have used in the present work a different
gauge for the electric field, a different method to derive the pair production
rate and a different representation for the physical operators $\hat{X}$ and
$\hat{P}$. However, this did not prevent us from obtaining the same results as
\cite{HK}.

\section{Hawking radiation in the presence of minimal length}

Interestingly, it is possible to make an immediate and clear link with some
interesting applications of particle creation in general relativity by the use
of simple arguments. The first application in this regard is the Unruh effect,
which concerns accelerated particle detectors in vacuum \cite{Holstein}.
Although all matter fields are in their vacuum states, the accelerated
detector will find a thermal distribution of particles with the Unruh
temperature%
\begin{equation}
T_{U}=\frac{a}{2\pi},
\end{equation}
where $a$ is the acceleration of the particle detector. Certainly, in
Schwinger effect, we don't have an accelerating detector. Instead, we have a
fixed detector and a particle of charge $e$ and mass $m$ subjected to a
constant electric field $E_{0}$. In such a case, the particle acceleration
$a=\frac{eE}{m}$ plays the same role as the detector acceleration and the
particle creation probability can be written as follows%

\begin{equation}
\mathcal{W}=\exp\left(  -\pi\frac{m^{2}}{eE}\right)  =\exp\left(  -2\pi
\frac{\mu}{a}\right)  ,
\end{equation}
where $\mu=\frac{m}{2}$ is the reduced mass of the produced pair. Therefore we
identify the Unruh temperature $T_{U}=\frac{m}{2\pi eE}=\frac{a}{2\pi}$.

The second interesting application concerns the black hole radiation due to
the particle creation. In black hole physics the situation is a bit different
since a static gravitational field without an event horizon can't create
particles. It was in the seventies when Hawking has, for the first time, shown
that a black hole can emit thermal spectrum with the Hawking temperature%
\begin{equation}
T_{H}=\frac{1}{8\pi GM},
\end{equation}
where $M$ is the black hole mass. We can outline a qualitative picture of the
Hawking radiation considering the scenario of creating virtual
particle-antiparticle pairs. Then one particle of the pair may happen to be
just outside of the black hole horizon while the other particle is inside it.
The particle inside the horizon inevitably falls onto the black hole center,
while the other particle can escape and may be detected by stationary
observers far from the black hole \cite{Win}.

Now, using the Unruh effect we can anticipate the Hawking result from the
constant electric field case by considering the gravitational acceleration at
the event horizon $r_{H}=2GM$ of the black hole%

\begin{equation}
a=\frac{GM}{r_{H}^{2}}=\frac{1}{4GM}.
\end{equation}
This reproduce the Hawking temperature from the Unruh one by replacing the
electric field acceleration $a=\frac{eE}{m}$ in Unruh's result by $a=\frac
{1}{4GM}$.

This correspondence between Schwinger effect and Hawking radiation can be,
straightforwardly generalized to the case where a minimal length exists. For
our case, by writing
\begin{equation}
\left(  1+\frac{1}{4}\beta m^{2}\left(  1-\frac{e^{2}E^{2}}{m^{4}}\right)
\right)  =\left(  1+\frac{1}{4}\beta m^{2}\right)  \left(  1-\frac{1}{4}%
\beta\frac{e^{2}E^{2}}{m^{2}}\right)
\end{equation}
we can see that%
\begin{equation}
\mathcal{W}\sim\exp\left(  -2\pi\frac{\tilde{\mu}}{a}\left(  1-\frac{\beta}%
{4}a^{2}\right)  \right)
\end{equation}
where $\tilde{\mu}=\frac{m}{2}\left(  1+\frac{1}{4}\beta m^{2}\right)  $. This
implies a modification on the Unruh temperature due to the minimal length%
\begin{equation}
\frac{1}{T_{U}^{GUP}}=\frac{2\pi}{a}\left(  1-\frac{\beta}{4}a^{2}\right)  .
\end{equation}
The correspondence between Unruh and Hawking effects, allows us to find
Hawking temperature corrected by the presence of a minimal length%
\begin{equation}
\frac{1}{T_{H}^{GUP}}=8\pi GM\left(  1-\frac{\beta}{64G^{2}}\frac{1}{M^{2}%
}\right)  .
\end{equation}
This is in good agreement with the corrected temperature obtained by
considering some scenarios which implies a generalized uncertainty principle
and preserve the relativistic dispersion relation (see for example
\cite{Amelino}). It should be noted that in some studies we remark that the
temperature correction is $4$ times greater than ours. This is because the
authors of these papers start from the uncertainty relation $\Delta P\Delta
X\geq\hbar\left(  1+\beta\left(  \Delta P\right)  ^{2}+...\right)  $ instead
of equation (\ref{eq:GUP}). The missing $\frac{1}{2}$ factor is the only
reason of this difference.

Furthermore, by the use of the first law of the black hole thermodynamics%
\begin{equation}
dM=TdS
\end{equation}
we obtain%
\begin{equation}
dS=\frac{dM}{T_{H}}=8\pi GM\left(  1-\frac{\beta}{64G^{2}}\frac{1}{M^{2}%
}\right)  dM,
\end{equation}
which leads to
\begin{equation}
S=4\pi GM^{2}-\pi\frac{\beta}{8G}\ln M+C \label{logterm}%
\end{equation}
where $C$ is a real constant and the logarithmic term is the well known
correction from quantum gravity to the classical Bekenstein--Hawking entropy,
which appears in different studies of GUP modified thermodynamics of black
holes. This result is, to the first order in $\beta$, exactly the same as that
recently communicated \cite{DuLong}
\begin{equation}
S_{A}=\frac{\pi\alpha}{16}\left[  \frac{2}{1-\sqrt{1-\frac{\alpha}{16}%
\frac{M_{p}^{2}}{M^{2}}}}+\ln\left(  1-\sqrt{1-\frac{\alpha}{16}\frac
{M_{p}^{2}}{M^{2}}}\right)  -\ln\left(  1+\sqrt{1-\frac{\alpha}{16}\frac
{M_{p}^{2}}{M^{2}}}\right)  \right]
\end{equation}
which can be developed to the first order in $\alpha$ to be%
\begin{equation}
S_{A}=4\pi\frac{M^{2}}{M_{p}^{2}}-\frac{\pi}{16}\alpha\ln\frac{M^{2}}%
{M_{p}^{2}}+...
\end{equation}
Here $M_{p}^{2}$ is the Planck mass $M_{p}=l_{P}^{-1}=\sqrt{\frac{1}{G}}$and
$\beta=\alpha M_{p}^{2}$.

Let us note that the corrected entropy for the Schwarzschild black hole can be
expressed as
\begin{equation}
S=\frac{A}{4G}-\frac{\pi}{16}\frac{\beta}{G}\ln\frac{A}{16\pi}, \label{CBHE}%
\end{equation}
where $A$ is the surface area of the black hole. This implies that the Hawking
radiation ceases before the black hole evaporates completely.

We also note that this finding can be generalized to many cases with the
spherically symmetric metric
\begin{equation}
ds^{2}=G\left(  r\right)  dt^{2}-\frac{1}{G\left(  r\right)  }dr^{2}%
+r^{2}d\Omega^{2}%
\end{equation}
where $G\left(  r\right)  $ is some real function that allows the existence of
an event Horizon. For this space-time metric we have the standard Hawking
temperature
\begin{equation}
T_{H}=\frac{\left\vert G^{\prime}\left(  r_{H}\right)  \right\vert }{4\pi}%
\end{equation}
where $r_{H}$ is the event horizon defined by $G\left(  r\right)  =0$. Then
according to the modification of the Unruh temperature we get
\begin{equation}
T_{H}^{GUP}=\frac{\left\vert G^{\prime}\left(  r_{H}\right)  \right\vert
}{4\pi}\left(  1+\frac{\beta}{16}\left\vert G^{\prime}\left(  r_{H}\right)
\right\vert ^{2}\right)  .
\end{equation}
As example we consider the case of the Rindler metric with $G\left(  r\right)
=1-2ar$. In this case we have $r_{H}=\frac{1}{2a}$ and $G^{\prime}\left(
r_{H}\right)  =-2a$. This gives%
\begin{equation}
T_{R}^{GUP}=\frac{a}{2\pi}\left(  1+\frac{\beta}{4}a^{2}\right)  .
\end{equation}
The second example is the static de-Sitter metric with $G\left(  r\right)
=1-H^{2}r^{2}$. For this metric, we have $r_{H}=\frac{1}{H}$ and $G^{\prime
}\left(  r_{H}\right)  =-2H^{2}r_{H}=-2H$. We then obtain%
\begin{equation}
T_{dS}^{GUP}=\frac{H}{2\pi}\left(  1+\frac{\beta}{4}H^{2}\right)  .
\end{equation}

\section{Exactness of the semiclassical computation}

As is mentioned above, it is remarkable that semi-classical computations,
using WKB method \cite{Semcla1} produce the exact amplitude for the
probability of pair production by constant electric field in ordinary
Minkowski space-time. In this section, we examine this WKB approximation in
the presence of a minimal length.

\subsection{Time dependent gauge}

Let us first consider the time dependent gauge, for which the pair creation
probability is, in the WKB\ approximation, given by \cite{Semcla2}
\begin{equation}
\mathcal{W}\approx\exp\left[  -2\operatorname{Im}\oint\sqrt{m^{2}+\left(
p+eEt\right)  ^{2}+\frac{2\beta}{3}\left(  p+eEt\right)  ^{4}}dt\right]  .
\end{equation}
By making the change $t\rightarrow z$, with%
\begin{equation}
z=\sqrt{eE}t+\frac{p}{\sqrt{eE}}%
\end{equation}
we can express $P$ as
\begin{equation}
\mathcal{W}\approx\exp\left[  -2\operatorname{Im}\int_{z_{0}^{\ast}}^{z_{0}%
}\sqrt{\lambda+z^{2}+2\alpha z^{4}}dz\right]  .
\end{equation}
where $z_{0}$ and $z_{0}^{\ast}$ are the turning points solutions of the
equation $\lambda+z^{2}+2\alpha z^{4}=0$.

Then by writing $\lambda+z^{2}+2\alpha z^{4}$ in the form
\begin{equation}
\lambda+z^{2}+2\alpha z^{4}=2\alpha\left(  z^{2}+b^{2}\right)  \left(
z^{2}+a^{2}\right)
\end{equation}
$\allowbreak$where$\allowbreak$%
\begin{align}
b^{2}  &  =-\frac{1}{4\alpha}+\frac{1}{4\alpha}\sqrt{1-8\alpha\lambda
}=-\left(  \lambda+2\alpha\lambda^{2}+8\alpha^{2}\lambda^{3}+...\right) \\
a^{2}  &  =-\frac{1}{4\alpha}-\frac{1}{4\alpha}\sqrt{1-8\alpha\lambda
}=-\left(  \frac{1}{2\alpha}-\lambda-2\alpha\lambda^{2}-8\alpha^{2}\lambda
^{3}+...\right)
\end{align}
with%
\begin{equation}
\frac{b}{a}=\sqrt{2\alpha\lambda}\left(  1+2\alpha\lambda+8\alpha^{2}%
\lambda^{2}\right)  .
\end{equation}
We can see that we have two pairs of turning points. In addition to the
expected two turning points that tend to the usual ones in the limit
$\beta\rightarrow0$, we have two extraturning points that will be rejected to
infinity in the limit $\beta\rightarrow0$. In this case we should studied the
contribution of each pair of turning points. It is also possible to have
interference between these turning points \cite{Dumlu}. To this aim we first
make the change $z\rightarrow ix$ and next we use of the following integrals
\begin{equation}
\int_{b}^{a}\sqrt{\left(  x^{2}-b^{2}\right)  \left(  x^{2}-a^{2}\right)
}dx=\frac{a}{3}\left[  \left(  a^{2}+b^{2}\right)  E\left(  \frac{b}%
{a}\right)  -2b^{2}K\left(  \frac{b}{a}\right)  \right]  ,
\end{equation}
and%
\begin{equation}
\int_{0}^{b}\sqrt{\left(  x^{2}-b^{2}\right)  \left(  x^{2}-a^{2}\right)
}dx=\frac{a}{3}\left[  \left(  a^{2}+b^{2}\right)  E\left(  \frac{b}%
{a}\right)  -\left(  a^{2}-b^{2}\right)  K\left(  \frac{b}{a}\right)  \right]
,
\end{equation}
where $K\left(  k\right)  $ and $E\left(  k\right)  $ are,respectively, the
Elliptic integrals of first and second kind which have the following
expansions for small values of $k$,%
\begin{align}
K\left(  k\right)   &  =\frac{\pi}{2}\left(  1+\frac{1}{4}k^{2}+\frac{9}%
{64}k^{4}+...\right) \\
E\left(  k\right)   &  =\frac{\pi}{2}\left(  1-\frac{1}{4}k^{2}-\frac{3}%
{64}k^{4}+...\right)  .
\end{align}
It is therefore easy to show that the contribution of the two ordinary turning
points
\begin{align}
\mathcal{W}_{1}  &  \approx\exp\left[  -4\sqrt{2\alpha}\int_{0}^{b}%
\sqrt{\left(  x^{2}-b^{2}\right)  \left(  x^{2}-a^{2}\right)  }dx\right]
\nonumber\\
&  =\exp\left[  -\pi\lambda\left(  1+\frac{3}{4}\alpha\lambda\right)  \right]
\end{align}
is very large compared to the contribution of the second pair of turning
points
\begin{equation}
\mathcal{W}_{2}\approx\exp\left(  -\frac{\pi}{\beta eE}\right)  .
\end{equation}
Thus, we obtain
\begin{equation}
\mathcal{W}\approx\exp\left[  -\pi\lambda\left(  1+\frac{3}{4}\alpha
\lambda\right)  \right]  =\exp\left[  -\pi\frac{m^{2}}{eE}\left(  1+\frac
{1}{4}\beta m^{2}\right)  \right]
\end{equation}
This shows that unlike the ordinary case, the semiclassical WKB approximation
does not give the exact results. The missing term is very important in the
strong field regime.

\subsection{Space dependent gauge}

In space dependent gauge, the study of particle creation reduces to the study
the tunneling effect. In the semiclassical approximation, it consists in
calculating the transmission coefficient $T$ which can be determined starting
from the Klein Gordon equation
\begin{equation}
\left[  \left(  \omega+eEx\right)  ^{2}+\frac{\partial^{2}}{\partial x^{2}%
}-\frac{2\beta}{3}\frac{\partial^{4}}{\partial x^{4}}-m^{2}\right]
\psi\left(  x\right)  =0, \label{KGx}%
\end{equation}
which is of fourth order. To the best of our knowledge, it does not admit
exact solutions. Nevertheless, we can obtain a more convenient equation by
introducing an auxiliary wave function $\varphi$, so that \cite{Salah}
\begin{equation}
\psi\left(  x\right)  =\left(  1+\frac{2\beta}{3}\frac{\partial^{2}}{\partial
x^{2}}\right)  \varphi\left(  x\right)  . \label{phi}%
\end{equation}
By substitute (\ref{phi}) into (\ref{KGx}) and neglecting terms of higher
order on $\beta$, we obtain
\begin{equation}
\left[  \left(  1+\frac{2\beta}{3}\left(  \left(  \omega+eEx\right)
^{2}-m^{2}\right)  \right)  \frac{\partial^{2}}{\partial x^{2}}+\left(
\omega+eEx\right)  ^{2}-m^{2}\right]  \varphi\left(  x\right)  =0,
\label{KGx2}%
\end{equation}
which is an effective Schr\"{o}dinger-like equation involving minimal length
corrections. In the limit $\beta\rightarrow0$, equation (\ref{KGx2}) reduces
to the ordinary Klein Gordon equation. Since this equation is of second order,
we can now apply the WKB method. First we write (\ref{KGx2}) in the form
\begin{equation}
\left[  \frac{\partial^{2}}{\partial x^{2}}+\tilde{p}^{2}\left(  x\right)
\right]  \psi\left(  x\right)  =0, \label{5.29}%
\end{equation}
with%
\[
\tilde{p}\left(  x\right)  =\frac{\left(  \omega+eEx\right)  ^{2}-m^{2}%
}{1+\frac{2\beta}{3}\left(  \left(  \omega+eEx\right)  ^{2}-m^{2}\right)  }%
\]
Using the WKB approximation, the probability of transition from a negative
energy state to a positive energy state is given by the transmission
coefficient \cite{Semcla3}
\begin{equation}
T=\exp\left(  -2\gamma\right)  \label{5.31}%
\end{equation}
where $\gamma$ is given by
\begin{equation}
\gamma=\int_{x_{1}}^{x_{2}}\left\vert p\left(  x\right)  \right\vert dx,
\label{5.32}%
\end{equation}
with%
\begin{equation}
\left\vert \tilde{p}\left(  x\right)  \right\vert =\frac{\sqrt{m^{2}-\left(
\omega+eEx\right)  ^{2}}}{1+\frac{\beta}{3}\left(  m^{2}-\left(
\omega+eEx\right)  ^{2}\right)  }. \label{5.33}%
\end{equation}
In (\ref{5.32}) $x_{1}$ and $x_{2}$ are the turning points solutions of the
equation $\tilde{p}\left(  x\right)  =0$. We have%
\begin{equation}
x_{1,2}=-\frac{\omega}{eE}\pm\frac{m}{eE} \label{5.34}%
\end{equation}
and%
\begin{align}
\gamma &  =\lambda\int_{-1}^{1}\sqrt{1-y^{2}}dy+\alpha\lambda^{2}\int_{-1}%
^{1}\left(  1-y^{2}\right)  \sqrt{1-y^{2}}dy.\label{5.35}\\
&  =\frac{\pi}{2}\lambda+\frac{3\pi}{8}\alpha\lambda^{2}%
\end{align}
Therefore, the factor $T$ takes the form
\begin{equation}
\mathcal{W}=T=\exp\left(  -\pi\lambda\left(  1+\frac{3}{4}\alpha
\lambda\right)  \right)  =\exp\left[  -\pi\frac{m^{2}}{eE}\left(  1+\frac
{1}{4}\beta m^{2}\right)  \right]  . \label{5.36}%
\end{equation}
Although different gauges are considered in this section, the final results
are the same. These results however are not exact. We have thus proven that
even at the semi-classical level the Schwinger effect preserves the gauge invariance.

\section{Conclusion}

In this paper we have studied the Schwinger effect, concerning the creation of
scalar particles by an electric field, in the presence of a minimal length.
The concept of minimum length is incorporated in the quantum mechanics of the
scalar particle through a specific generalization of the uncertainty principle
that necessitates a new definition of the momentum operator in position
representation. Consequently, the corresponding Klein Gordon equation contains
an inconvenient quartic term which impedes any attempt to exact solution.
However since the deformation parameter is so small, we were able to obtain an
approximate expression for the Green's function by introducing a
spacio-temporal transformation. Then by projecting this Green's function on
the outgoing particle and antiparticle states, we have calculated the pair
creation probability. Our results on the one hand confirm the results of
\cite{SCML} and on the other hand, suggests the consistency of the Schwinger
effect with the gauge invariance principle of electrodynamics.

The close analogy between Schwinger effect in a constant electric field and
the Unruh effect for accelerating observers enabled us to find the correction
brought by the minimal length on the Unruh temperature. This result, applied
to the Hawking radiation, has lead us to deduce the modified Hawking
temperature due to minimal length. Then from the first law of black hole
thermodynamics, we were able to extract the corrected black hole entropy. It
is shown that the first correction is a logarithmic term with a negative
numerical factor. This agrees with many anterior studies using different approaches.

As a last stage, we have examined the accuracy of the semiclassical WKB
approximation in the calculation of the pair production rate. Our results show
that, unlike the ordinary case, the WKB approximation in the presence of a
minimal length does not give the exact rate even for the constant electric field.

At the end of this work, let us make the following remarks:

\begin{itemize}
\item Semiclassical treatments are widely used in Schwinger effect and black
hole radiation. However, as we have seen in this work, it does not give the
exact results and one has to ask whether this can indeed be accurate in
computations of more complicated scenarios such as pair production by space or
time dependant electric fields or pair production in dynamical black holes
\cite{Firouzjaee}.

\item In addition, the position-momentum uncertainty relation considered in
this paper is not consistent with the usual Poincar\'{e} covariance, and
consequently, the corresponding minimal length is frame dependent. To deal
with this problem we have two possible paths; the first one is to consider,
like in \cite{MQ10}, a GUP model that is invariant under the usual
Poincar\'{e} group. The second is to search for a deformation of the
Poincar\'{e} group that makes the present minimal length invariant.
\end{itemize}

We leave these issues that demand more detailed investigations to future
works. Indeed the problem of particle creation in dynamical black holes with
minimal length is under consideration

\begin{acknowledgement}
M Bouali would like to thank L Cheriet for useful conversations. This work is
partially supported by Algerian Ministry of High Education and Scientific
Research and DGRSDT under the PRFU project: B00L02UN180120200001.
\end{acknowledgement}


\begin{thebibliography}{99}                                                                                               %


\bibitem {Sabine}S. Hossenfelder, Living Rev. Relativity 16 (2013) 2.

\bibitem {Tawfik}A. Tawfik and A. Diab, Int. J. Mod. Phys. D 23 (2014) 1430025

\bibitem {Mead}C. A. Mead, Phys. Rev. B 135, 849 (1964)

\bibitem {GUP1}M. Maggiore, Phys. Lett. B 319 (1993) 83.

\bibitem {GUP2}S. Hossenfelder, Mod. Phys. Lett. A 19 (2004) 2727.

\bibitem {GUP3}S. Hossenfelder, Phys. Rev. D 70 (2004) 105003.

\bibitem {GUP4}S. Hossenfelder, Phys. Lett. B 598 (2004) 92.

\bibitem {ST1}G. Veneziano, Europhys. Lett. 2 (1986) 199.

\bibitem {ST2}D. Amati, M. Ciafaloni, and G. Veneziano, Phys. Lett. B 197
(1987) 81

\bibitem {ST3}K. Konishi, G. Paffuti, and P. Provero, Phys. Lett. B 234 (1990) 276.

\bibitem {ST4}M. Kato, Phys. Lett. B 245 (1990) 43

\bibitem {LQG}L. J. Garay, Int. J. Mod. Phys. A 10 (1995) 145

\bibitem {BH1}F. Scardigli, Phys. Lett. B 452 (1999) 39.

\bibitem {BH2}F. Scardigli and R. Casadio, Class. Quant. Grav. 20 (2003) 3915.

\bibitem {BH3}R. J. Adler and D. I. Santiago, Mod. Phys. Lett. A 14 (1999) 1371.

\bibitem {NC1}M. R. Douglas and N. A. Nekrasov, Rev. Mod. Phys. 73 (2001) 977.

\bibitem {NC2}S. Minwalla, M. Van Raamsdonk and N. Seiberg, JHEP 02 (2000) 020.

\bibitem {NC3}R. J. Szabo, Phys. Rep. 378 (2003) 207.

\bibitem {Casadio}R. Casadio and F. Scardigli, Phys. Lett. B 807 (2020) 135558

\bibitem {Scardigli}F. Scardigli, R. Casadio, Eur. Phys. J. C 75 (2015) 425

\bibitem {MQ1}L. N. Chang, D. Minic, N. Okamura and T. Takeuchi, Phys. Rev. D
65 (2002) 125027.

\bibitem {MQ2}R. Akhoury and Y-P. Yao, Phys. Lett. B 572 (2003) 37.

\bibitem {MQ3}K. Nouicer, J. Phys. A: Math. Gen. 38 (2005) 10027.

\bibitem {MQ4}M. M. Stetsko and V. M. Tkachuk, Phys. Rev. A 74 (2006) 012101.

\bibitem {MQ5}D. Bouaziz and M. Bawin, Phys. Rev. A 78 (2008) 032110.

\bibitem {MQ6}D. Bouaziz and N. Ferkous, Phys. Rev. A 80 (2010).

\bibitem {MQ7}M. Asghari, P. Pedram and K. Nozari, Phys. Lett. B 725 (2013) 451

\bibitem {MQ8}A. Kempf, J. Math. Phys. 35 (1994) 4483.

\bibitem {MQ9}A. Kempf, G. Mangano and R. B. Mann, Phys. Rev. D 52 (1995) 1108.

\bibitem {Casimir1}K. Nouicer, J. Phys. A: Math. Gen. 39 (2006) 5125.

\bibitem {Casimir2}U. Harbach and S. Hossenfelder, Phys. Lett. B 632 (2006) 379.

\bibitem {Unruh1}P. Nicolini and M. Rinaldi, Phys. Lett. B 695 (2011) 303

\bibitem {BHT1}K Nouicer, Phys. Lett. B 646 (2007), 63-71

\bibitem {BHT2}R. V. Maluf and J. C. S. Neves, Phys. Rev. D 97, 104015 (2018)

\bibitem {BHT3}M. Cavaglia, S. Das, Class.Quant.Grav. 21 (2004) 4511-4522

\bibitem {BHT4}B. Majumder, Phys. Lett. B 703 (2011) 402-405

\bibitem {BHT5}A. Alonso-Serrano and M. Liska, Phys. Rev. D 104, (2021) 084043

\bibitem {BHT6}S. Das, P. Majumdar, R. K. Bhaduri, Class.Quant.Grav.19 (2002) 2355-2368,

\bibitem {HK}S. Haouat and K. Nouicer, Phys. Rev. D 89, (2014) 105030

\bibitem {SCML}F. Lu, B. Lv, P. Wang, and H. Yang, Nuclear Physics B 937
(2018) pp. 502--532.

\bibitem {BRMu}B-R. Mu, P. Wang, H.T. Yang, Commun. Theor. Phys. 63 (2015) 715

\bibitem {ANaqash}A. A. Naqash, B. Majumder, S. Mitra, M. M. Bangle, M.
Faizal, Eur. Phys. J. C 81 (2021) 870

\bibitem {EH}W. Heisenberg and H. Euler, Z. Phys. 98 (1936) 714.

\bibitem {Schwinger}J. Schwinger, Phys. Rev. 82 (1951) 664

\bibitem {Gelis}F. Gelis and N. Tanji, Prog. Part. Nucl. Phys. 87 (2016) 1.

\bibitem {Hossenfelder}S. Hossenfelder, Phys. Rev. D 73 (2006) 105013

\bibitem {Pad}T Padmanabhan, Pramana - J. Phys. 37, (1991) 179

\bibitem {Ong}Y.C. Ong, Eur. Phys. J. C 80, 777 (2020)

\bibitem {Hamil1}B. Hamil, M. Merad and T. Birkandan, Int. J. Mod. Phys. A 35
(2020) 2050014

\bibitem {Hamil2}B. Hamil and M. Merad, Int. J. Mod. Phys. A 33, 1850177 (2018).

\bibitem {NCPC1}N. Chair and M. Sheikh-Jabbari, Phys. Lett. B 504 (2001) 141

\bibitem {NCPC2}N. Mehdaoui, L. Khodja and S. Haouat, Int. J. Mod. Phys. A 36,
(2021) 2150011

\bibitem {Ruffini}R. Ruffini, G. Vereshchagin, S-S. Xue, Phys. Rep. 487 (2010)

\bibitem {CF1}S. P. Gavrilov, D. M. Gitman, Phys. Rev. D 53 (1996) 7162

\bibitem {Aitchison}I. J. R. Aitchison, Contemp. Phys. 26 (1985) 333-391

\bibitem {Cheriet}L. Cheriet and S Haouat, Ermakov-Penny equation and particle
creation: semi-classical treatement, To be submitted.

\bibitem {Grad}I. S. Gradshteyn and I. M. Ryzhik, Table of Integrals, Series
and Products (Academic Press, New York 1979)

\bibitem {EA1}G. Dunne and T. M. Hall, Phys. Lett. B\ 419, 322 (1998)

\bibitem {EA2}G. Dunne, G.V. Dunne, Heisenberg-Euler effective Lagrangians:
basics and extensions, in From fields to strings: circumnavigating theoretical
physics, M. Shifman et al. eds., World Scientific, Singapore (2005), p. 445.

\bibitem {EA3}S. Haouat and L. Chetouani, Phys. Scr. 75, 759 (2007).

\bibitem {EA4}S. P. Kim, H. K. Lee, Y. Yoon, Phys.Rev. D 78, 105013 (2008)

\bibitem {Ins1}G. Dunne and C. Schubert, Phys. Rev. D 72, 105004 (2005)

\bibitem {Ins2}G. Dunne, Q. Wang, H. Gies, and C. Schubert, Phys. Rev. D 73,
065028 (2006)

\bibitem {Ins3}S. P. Kim and D. N. Page, Phys.Rev. D 65, 105002 (2002)

\bibitem {Ins4}S. P. Kim and D. N. Page, Phys. Rev. D 73, 065020 (2006).

\bibitem {Grib}A. A. Grib, S. G. Mamayev and V.M. Mostepanenko,Vacuum Quantum
Effects in Strong Fields ( Friedmann Lab. Publ., St. Petersburg 1994)

\bibitem {Grib1}A. A. Grib, S. G.\ Mamayev and V.\ M.\ Mostepanenko, Gen. Rel.
Grav. 7 (1976) 535.

\bibitem {Che1}S. Haouat and R. Chekireb, Mod. Phys. Lett. A 26, (2011) 2639

\bibitem {Che2}S. Haouat and R. Chekireb, Int. J. Theor. Phys. 51 (2012) 1704

\bibitem {Che3}S. Haouat and R. Chekireb, Eur. Phys. J. C 72 (2012) 2034

\bibitem {Che4}S. Haouat and R. Chekireb, Int J Mod Phys A 30 (2015) 1550081

\bibitem {qka1}Y. Kluger, J. M. Eisenberg, B. Svetitsky, F. Cooper and E.
Mottola, Phys. Rev. Lett. 67 (1991) 2427.

\bibitem {qka2}D. B. Blaschke, A. V. Prozorkevich, G. Roepke, C. D. Roberts,
S. M. Schmidt, D. S. Shkirmanov, S. A. Smolyansky, Eur. Phys. J. D 55 (2009) 341

\bibitem {qka3}C. K. Dumlu, Phys. Rev. D 79 (2009) 065027

\bibitem {GF1}S. W. Hawking and J. B. Hartle, Phys. Rev. D 13 (1976) 2188

\bibitem {GF2}D. M. Chitre and J. B. Hartle, Phys. Rev D 16 (1977) 251.

\bibitem {GF3}A. O. Barut and I. H. Duru, Phys. Rev. D 41 (1990) 1312.

\bibitem {GF4}I. H. Duru and N. Unal, Phys. Rev. D 34 (1986) 959.

\bibitem {GF5}I. H. Duru, Gen. Rel. Grav. 26 (1994) 969.

\bibitem {GF6}H. Aoyama, M. Kobayashi, Prog. Theor. Phys. 64 (1980) 1045.

\bibitem {Semcla1}I. B. Khriplovich, Phys. Rep. 320 (1999) 37

\bibitem {Semcla2}V.S. Popov, JETP Lett. 13 (1971) 185.

\bibitem {Semcla3}S. Biswas, J. Guha and N. G. Sarkar; Class. Quantum Grav. 12
(1995) 1591

\bibitem {McLaughlin}D. McLaughlin and L. S. Schulman, J. Math. Phys. 12
(1971) 2520

\bibitem {Grosch}C. Grosch and F. Steiner, Handbook of Feynman Path Integrals,
Springer Tracts in Modern Physics 145 ( Springer, Berlin, Heidelberg 1998)

\bibitem {Holstein}B.R. Holstein, Am. J. Phys. 67 (1999) 499

\bibitem {Win}V. F. Mukhanov and S. Winitzki, Introduction to Quantum Effects
in Gravity. Cambridge University Press, Cambridge (2007)

\bibitem {Amelino}G. Amelino-Camelia, M. Arzano, Yi Ling and G. Mandanici,
Class. Quant. Grav 23 (2006) 2585

\bibitem {DuLong}X-D Du and C-Y. Long, JCAP 04 (2022) 031

\bibitem {Dumlu}C.K. Dumlu and G.V. Dunne, Phys. Rev. D 84 (2011) 125023

\bibitem {Salah}S. haouat, Phys. Lett. B 729 (2014) 33

\bibitem {MQ10}C. Quesne and V. M. Tkachuk, J. Phys. A 39 (2006) 10909

\bibitem {Firouzjaee}J. T. Firouzjaee, G. F. R. Ellis, Eur. Phys. J. C 76
(2016) 620
\end{thebibliography}
\end{document}